\def\o{\over}
\def\A{\rightarrow}
\def\bar{\overline}

\def\a{\alpha}
\def\b{\beta}
\def\n{\nu}
\def\m{\mu}

\def\e{\epsilon}

\def\th{\theta}

\def\bar{\overline}

\def\G{{\rm GeV}}

\def\eV{{\rm eV}}
\documentstyle [12pt]{article}
\begin{document}
\baselineskip=24.5pt
\setcounter{page}{1}
\thispagestyle{empty}
\topskip 0.5  cm
\begin{flushright}
\begin{tabular}{c c}
& {\normalsize   EHU-96-2}\\
& February 1996
\end{tabular}
\end{flushright}
\vspace{1 cm}
\centerline{\Large\bf  Structures of Neutrino Flavor Mixing  Matrix and}
\centerline{\Large\bf  Neutrino  Oscillations at CHORUS and NOMAD}
\vskip 1.5 cm
\centerline{{\bf Morimitsu TANIMOTO}
  \footnote{E-mail address: tanimoto@edserv.ed.ehime-u.ac.jp}}
\vskip 0.8 cm
 \centerline{ \it{Science Education Laboratory, Ehime University, 
 790 Matsuyama, JAPAN}}
\vskip 2.5 cm
\centerline{\bf ABSTRACT}\par
\vskip 0.5 cm
 We have studied structures of the neutrino flavor mixing  matrix focusing
 on the neutrino oscillations at CHORUS and NOMAD
 as well as  the one at LSND(or KARMEN).
We  have assumed two typical neutrino mass hierarchies 
  $m_3\simeq m_2\gg m_1$ and  $m_3\gg m_2\gg m_1({\rm or}\simeq m_1)$.
 Taking into account the see-saw mechanism of  neutrino masses,
   the reasonable neutrino flavor mixing patterns
have been  discussed.
The observation of the neutrino oscillation at
 CHORUS and NOMAD presents us the important constraint for
the structure of the neutrino flavor mixing matrix.
The atomospheric neutrino anomaly has been discussed
 in relation to the CHORUS and NOMAD experiments.
\par
\newpage
\topskip 0 cm
 \noindent 
{\bf 1. Introduction}\par
 Neutrino flavor oscillations are  important phenomena to search,
at low energy,   for physics beyond the Standard Model of the electroweak
 interaction, and to get information  on very high energy scales 
via the see-saw mechanism of the neutrino masses[1].
However, the only possible evidences for neutrino oscillations
 originate from the natural beams:
 the solar neutrinos[2] and the atomospheric neutrinos[3,4,5].
In the near future, data from the accelerator and reactor neutrino experiments
 will be available. CHORUS[6] is expected to present the first result soon.
 The first long base line reactor experiment CHOOZ[7]
will begin to operate.
 These experiments may resolve neutrino puzzles.
 \par
The tentative indication has been already given by the LSND experiment[8].
	 It was reported that an excess of 9 electron
events was observed at LSND.
If these events are due to the neutrino oscillation, the average 
$\bar \n_\m\A \bar \n_e$ oscillation
 transition probability is equal to
 $0.34^{+0.20}_{-0.18}\pm 0.07\%$[8,9].
KARMEN experiment[10]
 is also searching for the $\n_\m\A\n_e(\bar \n_\m\A \bar \n_e)$ oscillation 
 as well as LSND. The  CHORUS[6] and  NOMAD[11] experiments at CERN  are looking for
 the $\n_\m\A\n_\tau$ oscillation. 
The  most powerful reactor experiments searching for the neutrino oscillation
are Bugey[12]   and Krasnoyarsk[13] at present.
They provide excluded regions in  $(\sin^2 2\th, \Delta m^2)$
 parameter space by non-observation of the neutrino oscillation.\par
\par
 One expects to extract the neutrino flavor mixing matrix from 
  the data of neutrino flavor  oscillations.
In this paper, we study structures of the neutrino flavor mixing  matrix
 focusing  on the neutrino oscillations at CHORUS[6] and NOMAD[11]
 as well as the one at LSND(or KARMEN). 
 We find that the observation of the neutrino oscillation at
 CHORUS and NOMAD presents us the important constraint for
the structure of the neutrino flavor mixing matrix.
\par
  It is emphasized that there are only two hierarchical mass difference scales 
 $\Delta m^2$  in the three-flavor mixing scheme 
without introducing sterile neutrinos.
The neutrino with the mass $1\sim 7\eV$ is a candidate of
 the dark matter.  In particular, the cold+hot dark matter model
 has been shown to agree well with cosmological observations,
galaxies formation[14].
If we take this mass scale for the hightest one in the neutrino masses, 
 the other mass scale is
  either the atomospheric neutrino mass scale  $\Delta m^2\simeq 10^{-2}\eV^2$
 or the solar neutrino one $\Delta m^2\simeq 10^{-5}\sim
 10^{-6}\eV^2$.
 In our analyses, one neutrino mass scale is taken to be $1\sim 7\eV$,
  below which  CHORUS and NOMAD experiments will not be fruitful[6,11].
 The other scale is fixed  to be  the atomospheric neutrino one.
 The solar neutrino problem will be also discussed briefly in section 4.
By using the recent data from the accelerator and reactor neutrino
 experiments[15-18], we investigate the probability of the
 $\n_\m\A\n_\tau$ oscillation at CHORUS and NOMAD.
  Based on these results, 
we  discuss the structure of the neutrino flavor mixing matrix.\par
In section 2, we give  the formulation of the neutrino oscillations,
 and  we discuss  constraints from present accelerator and 
 reactor data in section 3.  In section 4, we present
numerical analyses at CHORUS and NOMAD,
 and discuss structures of the neutrino flavor mixing matrix. 
Section 5 is devoted to conclusions.
\vskip 0.2 cm
\noindent
{\bf 2. Formulation of Neutrino Oscillations}\par  
Recent  analyses of the three flavor neutrino oscillation
 are helpful for our work[19-23].
 In particular, the quantitative results by
 Bilenky et al.[20] and  Fogli, Lisi and  Scioscia[22]
 are the useful guide for our analyses although we
 concentrate mainly on the CHORUS and NOMAD results, which
 will be presented soon. \par
 The oscillation probability of neutrinos of energy $E_\n$
 after traversing the distance $L$ can be written as
\begin{equation}        
P(\n_\a\A \n_\b) = \delta_{\a\b} - 4 \sum_{i>j} 
               U_{\a i} U_{\b i} U_{\a j} U_{\b j} 
            \sin^2({\Delta m^2_{ij} L \o 4 E_\n}) \ , 
\end{equation}
\noindent  where  $\Delta m^2_{ij}=m^2_i-m^2_j$ is defined, and  
$U_{\a i}$ denote the elements of the $3\times 3$ neutrino  flavor 
mixing matrix, in which $\a$ and $i$  refer to the flavor eigenstate and 
 the mass eigenstate, respectively.  Since we neglect the $CP$ 
violation, the mixing parameters $U_{\a i}$ are real in our analyses.
We  assume   two typical hierarchical relations
    $\Delta m^2_{31}\gg \Delta m^2_{32}$
and $\Delta m^2_{31}\gg \Delta m^2_{21}$ in order to
 guarantee two different mass scales.
 The former relation corresponds to $m_3\simeq m_2\gg m_1$ 
 and the latter one to $m_3\gg m_2\gg m_1({\rm or}\simeq m_1)$.
 In our analyses, the highest neutrino mass scale is taken to be 
  $\Delta m^2_{31}=1\sim 50 \eV^2$, which is appropriate for the cosmological
 hot dark matter.\par
 The value of $L/E_\n$  is fixed in each experiment.
 In our following analyses, for example,  $L=800{\rm m}$ and $E_\n=30\G$ are 
taken   for the  $\n_\m\A\n_\tau$ experiment at CHORUS,
 and $L=30{\rm m}$ and $E_\n=36\sim 60\G$ 
  for the  $\bar \n_\m \A \bar\n_e$ experiment at LSND.
 For the atomospheric neutrino, $L=10\sim 10^4 {\rm km}$ and 
$E_\n=0.1\sim 10\G$  are expected.
  The approximate formulae of the oscillation probabilities are given 
 for  the accelerator and the reactor as follows.
For the case of $\Delta m^2_{31}\gg \Delta m^2_{32}$, those are given as

\begin{equation}        
P(\n_\m\A \n_e) =  4 U^2_{e 1} U^2_{\m 1} 
            \sin^2({\Delta m^2_{31} L \o 4 E_\n}) \ , 
\end{equation}
\begin{equation}        
P(\n_\m\A \n_\tau) =4 U^2_{\m 1} U^2_{\tau 1} 
            \sin^2({\Delta m^2_{31} L \o 4 E_\n}) \ ,
\end{equation}
\begin{equation}        
P(\n_\a\A \n_\a) =1-4 U^2_{\a 1}(1-U^2_{\a 1}) 
            \sin^2({\Delta m^2_{31} L \o 4 E_\n}) \ .
\end{equation}
\noindent
On the other hand, 
for the case of $\Delta m^2_{31}\gg \Delta m^2_{21}$, those are written as

\begin{equation}        
P(\n_\m\A \n_e) =  4 U^2_{e 3} U^2_{\m 3} 
            \sin^2({\Delta m^2_{31} L \o 4 E_\n}) \ , 
\end{equation}
\begin{equation}        
P(\n_\m\A \n_\tau) =4 U^2_{\m 3} U^2_{\tau 3} 
            \sin^2({\Delta m^2_{31} L \o 4 E_\n}) \ ,
\end{equation}
\begin{equation}        
P(\n_\a\A \n_\a) =1-4 U^2_{\a 3}(1-U^2_{\a 3}) 
            \sin^2({\Delta m^2_{31} L \o 4 E_\n}) \ .
\end{equation}
 If the atomospheric neutrino anomaly is attributed to the
 $\n_\m\A \n_\tau$ oscillation, the relevant formulae
 are given  instead of eqs.(3) and (6),
\begin{equation}        
P(\n_\m\A \n_\tau) =-4 U_{\m 2} U_{\tau 2} U_{\m 3} U_{\tau 3} 
            \sin^2({\Delta m^2_{32} L \o 4 E_\n})+
 2 U^2_{\m 1} U^2_{\tau 1}  \ ,
\end{equation}
\noindent and
\begin{equation}        
P(\n_\m\A \n_\tau) =-4 U_{\m 1} U_{\tau 1} U_{\m 2} U_{\tau 2} 
            \sin^2({\Delta m^2_{21} L \o 4 E_\n})+
 2 U^2_{\m 3} U^2_{\tau 3}  \ , 
\end{equation}
\noindent  respectively since $L/E_\n$ of the atomospheric neutrino
 is much larger than $L/E_\n$ of the acceleators and the reactors.
The atomospheric neutrino anomaly may be due to the $\n_\m \A \n_e$
oscillation. Then, instead of eqs.(2) and (5), the probabilities are
given as
\begin{equation}        
P(\n_\m\A \n_e) =-4 U_{e 2} U_{\m 2} U_{e 3} U_{\m 3} 
            \sin^2({\Delta m^2_{32} L \o 4 E_\n})+
 2 U^2_{e 1} U^2_{\m 1}  \ ,
\end{equation}
\noindent and
\begin{equation}        
P(\n_\m\A \n_e) =-4 U_{e 1} U_{\m 1} U_{e 2} U_{\m 2} 
            \sin^2({\Delta m^2_{21} L \o 4 E_\n})+
 2 U^2_{e 3} U^2_{\m 3}  \ ,
\end{equation}
\noindent respectively.
Eqs.(2)$\sim$ (11) are used in the following analyses.
\vskip 0.2 cm
\noindent
{\bf 3. Constraints from Present Accelerator and Reactor Data}\par
Let us begin with  discussing  constraints of  the accelerator 
and reactor disappearance experiments. 
Since no indications in favor of neutrino oscillations were found in 
these experiments, we only  get the  allowed regions 
in $(U^2_{\a i}, \Delta m^2_{31})$ parameter space.
The  recent  Bugey reactor experiment[12] and  the CDHS[15] and CCFR[16]
 accelerator experiments give the bounds for the neutrino
 mixing parameters at the fixed value of $\Delta m^2_{31}$.
 We follow the analyses given by Bilenky et al.[20].
\par
From eqs.(4) and (7) the mixing parameters can be expressed
 in terms of the oscillation probabilities as[20]
\begin{equation}        
 U_{\a i}^2={1\o 2}(1\pm \sqrt{1-B_{\n_\a\n_\a}}) \  , 
\end{equation}
\noindent with
\begin{equation}        
B_{\n_\a\n_\a}={1-P(\n_\a\A \n_\a)\o\sin^2({\Delta m^2_{31} L \o 4 E_\n})}\ ,  \end{equation}
\noindent where $\a=e, \m$ and  $i=1,3$.
Therefore, the parameters $U_{\a i}^2$ at the fixed value of 
$\Delta m^2_{31}$ should satisfy one of the following inequalities: 
\begin{equation}        
 U_{\a i}^2 \geq {1\o 2}(1 + \sqrt{1-B_{\n_\a\n_\a}})\equiv a_\a^{(+)} \ ,
    \qquad {\rm or} \qquad
 U_{\a i}^2 \leq {1\o 2}(1 - \sqrt{1-B_{\n_\a\n_\a}})\equiv a_\a^{(-)} \ . 
 \end{equation}
\noindent In Table 1, we show the values of $a_e^{(\pm)}$ and
$a_\m^{(\pm)}$, which were  obtained[20]
 from the negative results of the Bugey[12], CDHS[15] and CCFR[16]
 experiments, for the typical values of $\Delta m^2_{31}=1,4, 6, 30, 50 \eV^2$.
\begin{center}
\unitlength=0.7 cm
\begin{picture}(2.5,1.5)
\thicklines
\put(0,0){\framebox(3,1){\bf Table 1}}
\end{picture}
\end{center}
\par
From eq.(14) 
it is noticed there are three allowed regions of $U_{e i}^2$
 and $U_{\m i}^2$ as follows[20,22]:
\begin{eqnarray}        
 (A)\quad U_{e i}^2 \geq a_e^{(+)} \ , \qquad  U_{\m i}^2 \leq  a_\m^{(-)} \ ,
    \nonumber \\
 (B)\quad U_{e i}^2 \leq a_e^{(-)} \ , \qquad U_{\m i}^2 \leq  a_\m^{(-)} \ ,
  \\  
 (C) \quad U_{e i}^2 \leq a_e^{(-)} \ , \qquad U_{\m i}^2 \geq  a_\m^{(+)} \ ,
    \nonumber
\end{eqnarray}
\noindent
 where $i=1$ or $3$ corresponding to the neutrino mass hierarchies.
 In addition to these constraints,  we should take account of
 the constraints by the E531[17] and E776[18] experimental data.
 In some cases, these constraints become important.\par
\vskip 0.2 cm
\noindent
{\bf 4. Numerical Analyses at CHORUS and NOMAD}\par
  We study the  $\n_\m\A \n_\tau$ oscillation at CHORUS(NOMAD)
 under the constraints of other experiments. In particular, 
  we discuss it in relation to the $\n_\m\A \n_e$ oscillation
  at LSND(KARMEN) and the $\n_\m\A \n_X$ oscillation of
 the atomospheric neutrino.
We  assume  two hierarchical relations
    $\Delta m^2_{31}\gg \Delta m^2_{32}$
 and $\Delta m^2_{31}\gg \Delta m^2_{21}$, ie.,
 the  mass hierarchies such as 
  $m_3\simeq m_2\gg m_1$ and  $m_3\gg m_2\gg m_1({\rm or}\simeq m_1)$.
 The former hierarchy was suggested by investigating the hot dark matter 
 in ref.[14], where $m_3\simeq m_2\simeq 2.4 \eV$, ie.,
 $\Delta m^2_{31}\simeq 6\eV^2$ 
 is prefered. We call this case  hierarchy I.
 In the latter case, $m_3\leq 7\eV$  is expected
  as the candidate of the hot dark matter although this value depends on
 the Hubble parameter. The case is called hierarchy II.
 We analyze the neutrino oscillations for both hierarchies I and II
 in the cases (A), (B) and (C) given in eq.(15).\par
 Let us start with discussing the case (A),  in which we have
  $U_{e i}^2 \geq a_e^{(+)}$ and $U_{\m i}^2 \leq  a_\m^{(-)}$.
 For the hierarchy I, eqs.(2), (3), (4), (8), (10) are adopted
 for the relevant neutrino oscillations.
  The magnitude of the $\n_\m\A \n_e$ oscillation depends on   $U_{\m 1}^2$
 because  $U_{e 1}^2$ is close to 1.
 Since  $P(\n_\m\A \n_e)=P(\bar \n_\m\A \bar\n_e)$
 is guaranteed in the $CP$ conserved limit,
 we express the oscillation probability at LSND 
as $P_{\rm LSND}\equiv P(\n_\m\A \n_e)$,
 while $P_{\rm CHORUS}$ denotes $P(\n_\m\A \n_\tau)$ at CHORUS.
 Then,  we have the following relation between
$P_{\rm CHORUS}$ and $P_{\rm LSND}$:
\begin{equation}  
P_{\rm CHORUS}\simeq P_{\rm LSND} \times U_{\tau 1}^2 
\sin^2(\Delta m^2_{31}{L \o 4 E_\n})_{\rm CHORUS}/
 \sin^2(\Delta m^2_{31}{L\o 4 E_\n})_{\rm LSND}  \ .    
\end{equation}
\noindent 
If we take $\Delta m^2_{31}\simeq 6\eV^2$,
 we can estimate the upper bound of $P_{\rm CHORUS}$ by using the
  LSND result.
  Even if the reported LSND events[8] are due to the neutrino oscillations, 
 $P_{\rm LSND}$ should be  lower than $1.9\times 10^{-3}$ which
is derived from the upper bound of E776[18].
By using this bound, we obtain  $P_{\rm CHORUS}\leq 3\times 10^{-6}$,
 which is hopeless to be probed by CHORUS and NOMAD.\par 
 In the case of the hierarchy II, eqs.(5), (6), (7), (9), (11) are adopted
 for the neutrino oscillations.
 The mixing parameters $U_{\a 3}^2$ play an important role
 instead of $U_{\a 1}^2$.  The numerical discussion is completely
 parallel to the one in the hierarchy I.
 By using the E776 upper bound[18], 
 we get $P_{\rm CHORUS}\leq 3\times 10^{-6}(\Delta m^2_{31}= 6\eV^2)
 \sim 1.2\times 10^{-4}(\Delta m^2_{31}= 50\eV^2)$, which
 are out of the sensitivity at CHORUS and NOMAD.\par
The structures of the neutrino flavor mixing matrix ${\bf U}$ are
 different between the hierarchies  I and II. 
The mixing matrix in the hierarchies I is written as
\begin{equation}  
 {\bf U} \simeq \left (\matrix{ 1 &\e_3 &\e_4 \cr 
            \e_1 &U_{\m 2}&U_{\m 3}\cr
             \e_2 &U_{\tau 2}&U_{\tau 3} \cr} \right ) \ ,
\end{equation}
\noindent
 where  $\e_i(i=1\sim 4)$ are tiny numbers.
 As seen in eq.(8), the atomospheric neutrino anomaly could be
  solved by the large $\n_\m\A \n_\tau$ oscillation
by taking $\Delta m^2_{32}\simeq 0.01\eV^2$.
 On the other hand, 
it is impossible to explain the  anomaly
 by the large $\n_\m\A \n_e$ oscillation in eq.(10) because 
$U_{e 2}=\e_3$ and $U_{e 3}=\e_4$ are very small.
The  mixing matrix 
\begin{equation}  
 {\bf U} \simeq \left (\matrix{ 1 &\e_3 &\e_4 \cr 
            \e_1 &{1\o \sqrt{2}}&{1\o \sqrt{2}}\cr
             \e_2 &-{1\o \sqrt{2}}&{1\o \sqrt{2}} \cr} \right ) \ ,
\end{equation}
\noindent
is consistent with the atomospheric neutrino anomaly in Kamiokande[5]. 
Thus the hierarchy I could be consistent with the atomospheric neutrino anomaly
 although we cannot expect signals of the neutrino oscillation at CHORUS and NOMAD.

 For the hierarchy II, the mixing matrix is given as
\begin{equation}  
 {\bf U} \simeq \left (\matrix{ \e_1 &\e_2 &1 \cr 
            U_{\m 1} &U_{\m 2}& \e_3\cr
            U_{\tau 1} &U_{\tau 2}& \e_4 \cr} \right ) \ .
\end{equation}
\noindent
 The atomospheric neutrino anomaly
 is also explained by the large $\n_\m\A \n_\tau$ oscillation
with $\Delta m^2_{21}\simeq 0.01\eV^2$[5] in eq.(9).
 Then, the  mixing matrix is expected to be
\begin{equation}  
 {\bf U} \simeq \left (\matrix{ \e_1 &\e_2 &1 \cr 
            {1\o \sqrt{2}} & {1\o \sqrt{2}}& \e_3\cr
            -{1\o \sqrt{2}} &{1\o \sqrt{2}}& \e_4 \cr} \right ) \ .
\end{equation}
This  matrix structure in the hierarchy II
 mean that the heaviest neutrino is almost  the flavor $"e"$ neutrino.
 In this case, the neutrino mass hierarchy is a inverse one compared with
 the generation hierarchy of quark masses.
 The inverse mass hierarchy of $\n_3$(1st-generation) and 
$\n_1$(3rd-generation)  should be derived  from the structure of 
the right-handed
Majorana mass matrix in the see-saw mechanism of the neutrino  masses[1].
Since $\Delta m^2_{31}/\Delta m^2_{21}\simeq 10^{2\sim 4}$,
  the following relation  should be satisfied in order to guarantee 
   the inverse  neutrino masses, 
\begin{equation}  
  m_3\geq 10^{1\sim 2}\times m_1\ , \qquad {\rm with}
          \qquad m_3={m_1^{D2}\o M_1}\ , \qquad m_1={m_3^{D2}\o M_3} \ ,
\end{equation}
\noindent
where $m^D_1(m^D_3)$ and  $M_1(M_3)$ are the Dirac and  Majorana masses
 of the first(third) generation, respectively.  
Strictly speaking, this condition is not an exact one  because of
the off-diagonal elements in the mass matrices.  However,  we can
 safely neglect the contribution of the off diagonal elements since
the mixings  $U_{\m 3}$ and $U_{\tau 3}$ are very small.
 From eq.(21) we get 
\begin{equation}  
  {M_3\o M_1}\geq 10^{1\sim 2}\times \left ({m_3^D\o m_1^D}\right )^2\simeq 10^{10\sim 11} \ ,
\end{equation}
\noindent  where the t-quark  and  u-quark masses are taken
 for $m^D_3$ and $m^D_1$, respectively.
 This huge mass hierarchy of the Majorana masses seems to be 
 unnatural.
We conclude that the case (A) with the hierarchy II is ruled out
 because of the inverse hierarchy of the neutrino masses.\par
  In the case  (B),  we get $U_{\tau i}^2 \simeq 1$ since
  both $U_{e i}^2$ and  $U_{\m i}^2$ are very small as seen in eq.(15).
Then,  $P_{\rm CHORUS}$ depends on only $U_{\m i}^2$,
 which is constrained by the E531[17] bound of the $\n_\m\A \n_\tau$ 
 oscillation. On the other hand, $P_{\rm LSND}$ is suppressed
 because both $U_{e i}^2$ and  $U_{\m i}^2$ are very small.
In the  hierarchy I with $\Delta m^2_{31}\simeq 6\eV^2$,
 we get the bounds
\begin{equation}  
P_{\rm CHORUS} \leq 9.8\times 10^{-4} \ , \qquad\qquad
P_{\rm LSND} \leq 5.1\times 10^{-4} \ ,
\end{equation}
\noindent  where we used $U_{\m 1}^2\leq 6.0\times 10^{-3}$ at E531[17]
 and the  $U_{e 1}^2\leq 0.036$ at Bugey[12] in Table 1.
Obviously,  CHORUS and MOMAD  are expected to observe the neutrino oscillation
but our obtained $P_{\rm LSND}$ contradicts to the reported LSND probability,
 $0.34^{+0.20}_{-0.18}\pm 0.07\%$[8].
\par
In the  hierarchy II, $U_{e 3}^2$ and $U_{\mu 3}^2$ are relevant mixing
 parameters instead of $U_{e 1}^2$ and $U_{\mu 1}^2$.
These mixing parameters are also constrained 
              by the E531, CDHS and Bugey experiments. 
We show in fig.1 
the upper bounds of $P_{\rm CHORUS}$ and $P_{\rm LSND}$  with the 
limit of the sensitivity at CHORUS for $\Delta m^2_{31}=1\sim 50\eV^2$.
 The solid curve denotes the upper bound of  $P_{\rm CHORUS}$, which is
given by the E531 bound of $U_{\m 3}^2$[17] for $\Delta m^2_{31}=4\sim 50\eV^2$ and 
by the CDHS bound[15] for $\Delta m^2_{31}=1\sim 4 eV^2$.
The dashed curve denotes the upper bound of  $P_{\rm LSND}$, which is given by
   E531, CDHS and Bugey data.  As shown in fig.1, 
CHORUS and MOMAD  are hopeful to observe the neutrino oscillation,
however, our predicted upper bound of LSND is below
the recent reported LSND events[8].
\par
\begin{center}
\unitlength=0.7 cm
\begin{picture}(2.5,1.5)
\thicklines
\put(0,0){\framebox(3,1){\bf fig. 1}}
\end{picture}
\end{center}
\par
 As discussed in the case (A),  the structures of the neutrino flavor
mixing matrix ${\bf U}$ are also
 different between  the hierarchies I and II in the case (B). 
 The  mixing matrices being consistent with the atomospheric 
neutrino anomaly[5]  are expected to be
\begin{equation}  
 {\bf U} \simeq \left (\matrix{ \e_1 &{1\o \sqrt{2}}&{1\o \sqrt{2}}\cr 
            \e_2 &-{1\o \sqrt{2}}&{1\o \sqrt{2}}\cr
             1 & \e_3 & \e_4 \cr} \right ) \ ,
\end{equation}
\noindent
 for the hierarchy I, and
\begin{equation}  
 {\bf U} \simeq \left (\matrix{ {1\o \sqrt{2}} &{1\o \sqrt{2}}&\e_1\cr 
            -{1\o \sqrt{2}} &{1\o \sqrt{2}}&\e_2 \cr
             \e_3 & \e_4 & 1 \cr} \right ) \ ,
\end{equation}
 for the hierarchy II, respectively.
 It is noticed that , in the case (B), the atomospheric neutrino anomaly 
 could  occur due to the large $\n_\m\A \n_e$ oscillation
 as seen in eqs.(10) and (11).
 In the hierarchy I, the generation hierarchy of neutrino masses 
is the inverse one since
 $"\tau"$ neutrino is the lightest one, 
while  it is an ordinary one in the hierarchy II.
 The inverse neutrino mass hierarchy  leads to the unnatural
  huge mass difference of the Majorana masses 
 $O(10^{10\sim 11})$, as discussed in the case (A).
Therefore, the case (B) with the hierarchy I is  ruled out. \par
 One may consider that the case (B) with hierarchy II is a natural case consistent 
 with the atomospheric neutrino anomaly.  However,  the large $\n_\m - \n_e$ mixing
  is excluded by the reactor experiments at Bugey[12]  and Krasnoyarsk[13].
  The mixing angle is constrained such as
    $sin^2 2\th_{e\m}\leq 0.7$ in the case of $\Delta m^2_{21}=10^{-2}\eV^2$
	 in  the reactor experiments while the data of the atomospheric neutrino anomaly
 in Kamiokande[5] suggests $\Delta m^2_{21}=7\times 10^{-3}\sim 8\times 10^{-2} \eV^2$  
  and $sin^2 2\th_{e\m}= 0.6\sim 1$ for the  $\n_\m\A \n_e$ oscillation.
  The  overlap region is rather small such as 
  $sin^2 2\th_{e\m}= 0.6\sim 0.7$. The new reactor experiments will  soon give
  the  severer constraint for the $\n_\m -\n_e$ mixing.\par
   
  The case  (C) is in the region of $U_{\m i}^2 \simeq 1$. 
Then,  $P_{\rm CHORUS}$ depends on only $U_{\tau i}^2$,
 which is constrained by the E531[17] bound of the $\n_\m\A \n_\tau$ 
 oscillation. On the other hand, $P_{\rm LSND}$ depends on
  $U_{e i}^2$, which is constrained by the E776[18] bound of 
the $\n_\m\A \n_e$  oscillation.
Hence, there is no relation between $P_{\rm CHORUS}$ and $P_{\rm LSND}$
 for the present.
In the  hierarchy I with $\Delta m^2_{13}\simeq 6\eV^2$,
 we get the bounds
\begin{equation}  
P_{\rm CHORUS} \leq 9.8\times 10^{-4} \ , \qquad\qquad
P_{\rm LSND} \leq 1.9\times 10^{-3} \ ,
\end{equation}
\noindent  where we used the E531 bound $U_{\tau 1}^2\leq 6.0\times 10^{-3}$
 and the  E776 bound $U_{e 1}^2\leq 5.2\times 10^{-4}$.
 CHORUS and MOMAD  are expected to observe the neutrino oscillation.
Of course, the reported LSND events[8] are still consistent with our result.
\par
In the the hierarchy II, $U_{e 3}^2$ and $U_{\tau 3}^2$ are relevant mixing
 parameters instead of $U_{e 1}^2$ and $U_{\tau 1}^2$.
These parameters are bounded by the E531, CDHS and E776 experiments.
 We show  
the upper bounds of $P_{\rm CHORUS}$ and $P_{\rm LSND}$ in fig.2, in which
 notations are same as  in fig.1.  It is found that
 CHORUS and MOMAD  are also hopeful to observe the neutrino oscillation.
The prediction is also consistent with the reported LSND events[8].
\begin{center}
\unitlength=0.7 cm
\begin{picture}(2.5,1.5)
\thicklines
\put(0,0){\framebox(3,1){\bf fig. 2}}
\end{picture}
\end{center}
 The structures of the neutrino mixing matrix  are given as
 \begin{equation}  
 {\bf U} \simeq \left (\matrix{ \e_1 & U_{e 2} &U_{e 3} \cr 
            1 & \e_2 & \e_3 \cr
             \e_4 &U_{\tau 2}&U_{\tau 3} \cr} \right ) \ ,
\end{equation}
\noindent
 for the hierarchy I,  and  
 \begin{equation}  
 {\bf U} \simeq \left (\matrix{ U_{e 1} & U_{e 2} & \e_1 \cr 
            \e_2 & \e_3 & 1 \cr
             U_{\tau 1} &U_{\tau 2}&\e_4 \cr} \right ) \ ,
\end{equation}
\noindent
 for the hierarchy II. 
 In both hierarchies, we cannot explain
the atomospheric neutrino anomaly by the neutrino oscillations
 because $\e_2$ and $\e_3$ remain to be small due to the unitarity.
 Since  the flavor $"\m"$ couples to the lightest neutrino strongly
in the hierarchy I,
   the masses of $\n_1$ and $\n_2$  are  inverse  compared 
with  the generation of quarks.
On the other hand, the masses of  $\n_2$ and $\n_3$ are inverse in the hierarchy II
 as the flavor $"\m"$ couples to the heaviest neutrino.
 The magnitude of the Majorana mass hierarchy
 is estimated in the same way as in eqs.(21) and (22)
 by using the top-quark, c-quark and u-quark masses
for $m^D_3$, $m^D_2$ and $m^D_1$, respectively.
 The mass hierarchy of $M_1\sim M_3$ is roughly larger than $O(10^6)$,
  which may be allowed   to  build    a natural  model.\par
  
  We summarize  in Table 2 the predicted  upper bound of $P_{\rm CHORUS}$ and $P_{\rm LSND}$
   in the case of $\Delta m^2_{31}=6 \eV^2$, the  situation of the atomospheric neutrino anomaly,
    and the neutrino mass hierarchy for each case. 
\begin{center}
\unitlength=0.7 cm
\begin{picture}(2.5,1.5)
\thicklines
\put(0,0){\framebox(3,1){\bf Table 2}}
\end{picture}
\end{center}
\par
As the results of our analyses of cases (A), (B) and (C), 
we give  remarks on the atomospheric neutrino anomaly.
The anomaly
 could be explained by the $\n_\m$ oscillation in the cases (A) and (B).
 It is emphasized that 
 either $\n_\m\A \n_\tau$ or $\n_\m\A \n_e$ modes could  be operative
 in the atomospheric neutrino.
  There is no solution that both modes are dominant.
In our analyses, we fixed the second mass scale as the
 atomospheric neutrino mass scale. Therefore, we cannot explain
 the solar neutrino problem by the neutrino oscillation
without introducing sterile neutrino[24,25].
On the other hand, if we abandon the possibility of solving
 the atomospheric neutrino anomaly by the neutrino oscillation,
 we can take the second mass scale as the solar neutrino mass scale.
Taking the MSW solution[26], the solar neutrino problem could be
 reconciled with our analyses except for one case,  which
 is the case (A) with the hierarchy II as pointed out by Bilenky et al.[20].
   Since we have $U_{e3}\simeq 1$ in this exceptional case,
 the survival probability of the solar neutrinos is too large
 to be consistent with the data of GALLEX and SAGE, 
which have shown less neutrino deficit than the
Homestake and Kamiokande experiments[2].
\vskip 0.2 cm
\noindent
{\bf 5. Conclusions}\par
 We have studied structures of the neutrino flavor mixing  matrix focusing
 on the neutrino oscillations at CHORUS and NOMAD
 as well as the one at LSND(or KARMEN).
We  assumed the typical mass hierarchies 
  $m_3\simeq m_2\gg m_1$(hierarchy I) and  
$m_3\gg m_2\gg m_1({\rm or}\simeq m_1)$(hierarchy II).
 Taking into account the see-saw mechanism[1] of neutrino masses, 
the reasonable neutrino flavor mixing patterns have been  discussed. 
In the case (A), only the hierarchy I
 gives the reasonable neutrino flavor mixing matrix. Then, CHORUS and NOMAD
 are hopeless to observe the  neutrino oscillations although LSND(KARMEN) could
 observe the oscillation.  The atomospheric neutrino anomaly
 could be explained due to the large $\n_\m\A \n_\tau$ oscillation.
In the  case (B),  the hierarchy II
 gives the reasonable neutrino flavor mxing matrix.  CHORUS and NOMAD
 are hopeful to observe the evidence of the neutrino oscillations, but
 the expected upper bound of the LSND(KARMEN) experiment is much smaller than
 the present experimental sensitivity.
  We have a small chance to explain  the atomospheric neutrino anomaly
  by the large $\n_\m - \n_e$ mixing $sin^2 2\th_{e\m}= 0.6\sim 0.7$.
   The new reactor experiments will  soon  check this case.
 Although the  case (C) leads to  the inverse hierarchy  for the Majorana masses, 
  it may be still natural to  build a model. 
In this case,   both CHORUS(NOMAD) and LSND(KARMEN)
  could observe the evidence of the neutrino oscillations, but
the atomospheric neutrino anomaly
 could not be explained due to the neutrino oscillations.
\par
If CHORUS will observe the signal of the neutrino oscillation,
 the case (B) with the hierarchy II  and the case (C) 
  are  the  reasonable cases as seen  in Table 2.
Then, we have  to wait for more sensitive experiments
 at LSND or KARMEN in order to determine
 the value of the heaviest neutrino mass.
 The experimental study of the atomospheric neutrino anomaly
  in Super-Kamiokande is also important to decide the favorable case.
Thus, the observation of the neutrino oscillation at
 CHORUS and NOMAD presents us the important constraint for
the structure of the neutrino flavor mixing matrix,
 which is very useful guide to go beyond the Standard Model
 of the quark-lepton unification.\par
 
 What can we learn if CHORUS and NOMAD will observe  no signals of the neutrino oscillation?
 The case (A) with hierarchy I is most reasonable case 
  if the atomospheric neutrino anomaly is caused by the large flavor mixing. 
  Otherwise, we will need  more detailed studies  by using
	 the improved  limits of $U_{\m i}$  and $U_{\tau i}(i=1,3)$ 
	 by CHORUS and NOMAD in the case (C). 
\par
\vskip 0.3 cm
\centerline{\bf Acknowledgments}
I would like to thank Prof. A.Yu. Smirnov and Prof. H. Minakata 
for helpful discussions.
This research is supported by the Grant-in-Aid for Science Research,
Ministry of Education, Science and Culture, Japan(No. 07640413).
\newpage
 
\topskip 1 cm \centerline{{\bf References}}
\vskip 1 cm
\noindent
[1] M. Gell-Mann, P. Ramond and R. Slansky, in {\it Supergravity}, 
 Proceedings of the \par
Workshop, Stony Brook, New York, 1979, edited by P. van Nieuwenhuizen \par
and D. Freedmann, North-Holland, Amsterdam, 1979, p.315;\par
 T. Yanagida, in {\it Proceedings of the Workshop on the Unified Theories and 
Baryon\par
 Number in the Universe}, Tsukuba, Japan, 1979, edited by O. Sawada and \par
A. Sugamoto, KEK Report No. 79-18,
    Tsukuba, 1979, p.95.\par
\noindent
[2]  GALLEX Collaboration,  Phys. Lett.  {\bf 327B}(1994)377;\par
 SAGE Collaboration, Phys. Lett.  {\bf 328B}(1994)234;\par
 Homestake Collaboration, Nucl. Phys. {\bf B38}(Proc. Suppl.)(1995)47;\par
Kamiokande Collaboration, Nucl. Phys. {\bf B38}(Proc. Suppl.)(1995)55.
\par\noindent
[3] K.S. Hirata et al., Phys. Lett. {\bf 205B}(1988)416;
 {\bf 280B}(1992)146;\par
D. Casper et al., Phys. Rev. Lett. {\bf 66}(1991)2561;\par
 R. Becker-Szendy et al., Phys. Rev. {\bf D46}(1992)3720.\par
\noindent
[4]  NUSEX Collaboration, Europhys. Lett. {\bf 8}(1989)611;
  ibidem {\bf 15}(1991)559;\par
 SOUDAN2 Collaboration, Nucl. Phys. {\bf B35}(Proc. Suppl.)(1994)427;
\par  ibidem {\bf 38}(1995)337;\par
 Fr\'ejus Collaboration, Z. Phys. {\bf C66}(1995)417;\par
  MACRO Collaboration,  Phys. Lett.  {\bf 357B}(1995)481.
\par\noindent
[5] Y. Fukuda et al., Phys. Lett. {\bf 335B}(1994)237.\par\noindent
[6] K. Winter, Nucl. Phys. {\bf B38}(Proc. Suppl.)(1995)211;\par
    M. Baldo-Ceolin, ibidem {\bf 35}(1994)450.\par
\noindent
[7] R.I. Steinberg, Proc. of the 5-th Int.Workshop on neutrino Telescopes,
  Venice, \par
  Italy, ed. by M. Baldo-Ceolin(1993)209.
\par\noindent
[8] LSND Collaboration, C. Athanassopoulos et al., 
  Phys. Rev. Lett. {\bf 75}(1995)2650. 
  \par\noindent
[9] J. E. Hill, Phys. Rev. Lett. {\bf 75}(1995)2654.
\par\noindent
[10] KARMEN Collaboration, Nucl. Phys. {\bf B38}(Proc. Suppl.)(1995)235.
\par\noindent
[11] L. DiLella, Nucl. Phys. {\bf B31}(Proc. Suppl.)(1993)319.\par
\noindent
[12] B. Achkar et al., Nucl. Phys. {\bf B434}(1995)503.\par
\noindent
[13] G.S. Vidyakin et al., Pis'ma Zh.Eksp.Thor.Fiz.{\bf 59}(1994)364;
 \par JETP Lett. {\bf 59}(1994)390. \par\noindent
[14] J.R. Primack, J. Holtzman, A. Klypin
 and D. O. Caldwell, Phys. Rev. Lett. \par
 {\bf 74}(1995)2160. \par
\noindent
[15] CDHS Collaboration, F. Dydak et al., Phys. Lett. {\bf 134B}(1984)281.
 \par\noindent
[16] CCFR Collaboration, I.E. Stockdale et al., Phys. Rev. Lett.
     {\bf 52}(1984)1384;\par
     Z. Phys. {\bf C27}(1985)53.
 \par\noindent
[17] E531 Collaboration, N. Ushida et al., Phys. Rev. Lett. {\bf 57}(1986)2897.
 \par\noindent
[18] E776 Collaboration, L. Borodovsky et al., Phys. Rev. Lett. 
 {\bf 68}(1992)274. \par\noindent
[19] H. Minakata, Phys. Lett. {\bf 356B}(1995)61;\par
   preprint TMUP-HEL-9502, hep-ph/9503417.
\par\noindent
[20] S.M. Bilenky, A. Bottino, C. Giunti and C. W. Kim,
 Phys. Lett. {\bf 356B}(1995)273.\par
\noindent
[21] S.M. Bilenky,  C. Giunti and C. W. Kim,
 preprint DFTT 30/95, JHU-TIPAC 95017,\par
  hep-ph/9505301(1995).\par
\noindent
[22] G.L. Fogli, E. Lisi and G. Scioscia,
  preprint IASSNS-AST 95/28, \par
  BARI-TH/210-95(1995).
\par\noindent
[23] S. Goswami, K. Kar and A. Raychaudhuri
  preprint CUPP-95/3(1995).
\par\noindent
[24] E.J. Chun, A.S. Joshipura and A.Yu. Smirnov,
       Phys. Lett. {\bf 357B}(1995)608;\par  
preprint IC/95/76, PRL-TH/95-7, hep-ph/9505275(1995). \par\noindent
[25]  A.Yu. Smirnov,
  preprint at Yukawa Institute, YITP-95-3(1995). 
\par\noindent
[26] S.P. Mikheyev and A.Yu. Smirnov, Yad. Fiz.
     {\bf 42}(1985)1441;\par
  L. Wolfenstein, Phys. Rev. {\bf  D17}(1987)2369;\par
 E.W. Kolb, M.S. Turner and T.P. Walker,
   Phys. Lett. {\bf 175B}(1986)478;\par
 S.P. Rosen and J.M. Gelb, Phys. Rev.
         {\bf D34}(1986)969;\par
   J.N. Bahcall and H.A. Bethe, Phys. Rev. Lett.
     {\bf  65}(1990)2233;\par
N. Hata and P. Langacker, Phys. Rev. {\bf  D50}(1994)632;\par
 P.I. Krastev and A. Yu. Smirnov, Phys. Lett. {\bf 338B}(1994)282.
\par
\newpage
\topskip 2 cm
\centerline{\bf Figure Captions}
\par
\vskip 1 cm
\noindent
{\bf figure 1:}\par
 Upper bounds of $P_{\rm CHORUS}$ and $P_{\rm LSND}$ 
 versus $\Delta m^2_{31}$ in the case (B) with the hierarchy II.
 The solid and dashed curves correspond to  
 $P_{\rm CHORUS}$ and $P_{\rm LSND}$, respectively.
 The horizontal solid line denotes the limit of the sensitivity at CHORUS
 and the horizontal dashed one denotes the lower bound from the
 recent reported LSND events.
\par

\vskip 1 cm
\noindent
{\bf figure 2:}\par
 Upper bounds of $P_{\rm CHORUS}$ and $P_{\rm LSND}$ 
 versus $\Delta m^2_{31}$ in the case (C) with the hierarchy II.
Notations are same as in  fig.1.
\par
\newpage
\topskip 1 cm
\begin{table}
\hskip 3.7 cm
\begin{tabular}{| r| r| r| r| r|} \hline
                     &         &         &          &          \\
  $\Delta m^2(\eV^2)$ & $a_e^{(-)}\ \ $ & $a_e^{(+)}\ \ $ & $a_\m^{(-)}\ \ $ &
  $a_\m^{(+)}\ \ $  \\
                      &         &         &          &  \\ \hline
                      &         &         &          &        \\
  $1\qquad $           &0.011    & 0.989   & 0.028    & 0.972  \\
  $4\qquad $           &0.042    & 0.958   & 0.015    & 0.985  \\
  $6\qquad $           &0.036    & 0.964   & 0.020    & 0.980  \\
  $30\qquad$           &0.039    & 0.961   & 0.036    & 0.964  \\
  $50\qquad$           &0.039    & 0.961   & 0.028    & 0.972  \\
                      &         &         &          &     \\  \hline
\end{tabular}
\caption{ Values of the upper and lower bounds $a_\a^{(-)}$ and $a_\a^{(+)}$
($\a=e,\m$) for $U_{\a i}^2$}
\end{table}
\vskip 1.5 cm

\begin{table}
\hskip 0.5 cm
\begin{tabular}{| r| r| r| r| } \hline
                     &         &         &             \\
                     & (A) \ \ \ \ \ \ \ \   & (B) \ \ \ \ \ \ \ \ & (C)   \ \ \ \ \ \  \\
                     &         &         &            \\   \hline
                     &         &         &             \\
 I ($\Delta m^2_{31}=6 \eV^2$) \quad $P_{\rm CHORUS}$ 
               & $3\times 10^{-6}$ & $9.8\times 10^{-4}$ & $9.8\times 10^{-4}$ \\
                $P_{\rm LSND}\ \ \ $ 
        & $1.9\times 10^{-3}$ & $5.1\times 10^{-4}$ & $1.9\times 10^{-3}$    \\  
	Atomospheric Neutrino  & $\n_\m$-$\n_\tau$\ \ \ \ &	$\n_\m$-$\n_e$\ \ \ \  & no \ \ \ \ \ \ \\
 Neutrino mass hierarchy &ordinary \ \ \ & $\n_1$-$\n_3$ inverse $\times$ & $\n_1$-$\n_2$ inverse\\
              &            &             &              \\   \hline
            &                     &                       &     \\  
 II ($\Delta m^2_{31}=6 \eV^2$) \quad $P_{\rm CHORUS}$
              & $3\times 10^{-6}$ & $9.8\times 10^{-4}$ & $9.8\times 10^{-4}$ \\
      \qquad $P_{\rm LSND}\ \ \  $ 
		     & $1.9\times 10^{-3}$   & $5.1\times 10^{-4}$  &  $1.9\times 10^{-3}$     \\
 Atomospheric Neutrino & $\n_\m$-$\n_\tau$\ \ \ \ &	$\n_\m$-$\n_e$ \ \ \ \ & no \ \ \ \ \ \  \\
 Neutrino Mass Hierarchy & $\n_1$-$\n_3$ inverse $\times$ & ordinary\ \ \ & $\n_2$-$\n_3$ inverse\\
            &         &          &             \\  \hline
\end{tabular}
\caption{ Predicted upper bounds of $P_{\rm CHORUS}$ and  $P_{\rm LSND}$  
 in the case of $\Delta m^2_{31}=6 \eV^2$, the situation of the atomospheric neutrino anomaly,
 and the neutrino mass hierarchy for each case. 
   Here, $\times$ denotes the unnatural huge  Majorana mass hierarchy. }
\end{table}
 
\end{document}